# Assessing the Integration of Software Agents and Industrial Automation Systems with ISO/IEC 25010

Stamatis Karnouskos[1], Roopak Sinha[2], Paulo Leitão[3], Luis Ribeiro[4], Thomas. I. Strasser[5]

[1]*SAP, Walldorf, Germany, email: stamatis.karnouskos@sap.com*
[2]*IT & Software Engineering, Auckland University of Technology, New Zealand, email: roopak.sinha@aut.ac.nz*
[3]*Research Centre in Digitalization and Intelligent Robotics (CeDRI), Instituto Politécnico de Bragança,
Campus de Santa Apolónia, 5300-253 Bragança, Portugal, email: pleitao@ipb.pt*
[4]*Linköping University, SE-58 183 Linköping, Sweden, email: luis.ribeiro@liu.se*
[5]*Center for Energy – AIT Austrian Institute of Technology and Institute of Mechanics and Mechatronics –
Vienna University of Technology, Vienna, Austria, email: thomas.i.strasser@ieee.org*

**Abstract**

*Agent-technologies have been used for higher-level decision making in addition to carrying out lower-level automation and control functions in industrial systems. Recent research has identified a number of architectural patterns for the use of agents in industrial automation systems but these practices vary in several ways, including how closely agents are coupled with physical systems and their control functions. Such practices may play a pivotal role in the Cyber-Physical System integration and interaction. Hence, there is a clear need for a common set of criteria for assessing available practices and identifying a best- fit practice for a given industrial use case. Unfortunately, no such common criteria exist currently. This work proposes an assessment criteria approach as well as a methodology to enable the use case based selection of a best practice for integrating agents and industrial systems. The software product quality model proposed by the ISO/IEC 25010 family of standards is used as starting point and is put in the industrial automation context. Subsequently, the proposed methodology is applied, and a survey of experts in the domain is carried out, in order to reveal some insights on the key characteristics of the subject matter.*

## 1. INTRODUCTION

Industrial Agents (IA) have been used to introduce intelligence and adaptation in complex and dynamic industrial systems. Many of these *Cyber-Physical Systems* (CPS) containing distributed hardware and software, controlling complex physical processes, are also constrained by requirements such as the need to integrate low-level automation functions [1] . Such services typically include the interface of software agents to the underlying physical process (i.e., process interface) by using analog and digital I/Os, the handing of control programs (start, stop, update, etc.), as well as the execution of low-level automation and control algorithms [1] , [2] .

Agent approaches have been successfully used in a variety of domains like factory automation, power & energy systems, and building automation [3] . A recent detailed survey has identified and analyzed a range of such practices [2] . Within the scope of agent-oriented practices for industrial automation systems, the *IEEE P2660.1* working group [4] aims to provide some clarity, with a goal to promote recommended best practices for using industrial agents in order to enable the reuse and transparency. A key step in the process to pursue of the definition of a "best practice" is to define criteria that can be used to compare the strengths and limitations of available practices. While industry adoption of practices integrating IA and low-level automation functions is interlinked with several complex factors [5] , we specifically target the specification of a set of criteria that can be utilized in a clear and coherent manner for assessing existing practices within this work [2] .

A major challenge in proposing a set of criteria for the assessment and comparison of practices is the lack of a coherent and widely-accepted quality model. This challenge is discussed in section 2 where relevant industrial standards are identified that should be considered. This discussion helps in identifying and adopting the product quality model proposed by the ISO/IEC 25010 standards family [6] . Although this model is targeted towards software systems, it is a good fit for CPS. Furthermore its aspects are put in this work in the industrial automation context as discussed in section 4 . We also propose a methodology to guide the choice and/or assessment of practices integrating IA and low-level automation, that is presented in section 3. The applicability and limitations of these approaches are discussed in section 5 followed by the conclusions in section 6 .

This work focuses on the interface between software agents and physical industrial automation systems, and not the intelligent behavior of agents or the intercommunication among the agents. Overall the contributions are: (i) an investigation of the applicability of ISO/IEC 25010 characteristics and sub-characteristics for the practices integrating IA and low-level automation functions, (ii) a methodology for ranking or selecting an appropriate practice based on the desired qualities, which aims to help industry stakeholders in deciding which practices to use, and (iii) an expert survey revealing some insights on ISO/IEC 25010 characteristics for industrial automation systems.

## 2. RELATED WORK

Software complexity increases, and as such there have been significant efforts to be able to select approaches based on quality characteristics [7] . A comprehensive assessment would include functional, interface, performance, and physical requirements, as well as other non-functional requirements expressing the levels of safety, security, reliability etc. [8] . Capturing and specifying these requirements, especially in a measurable manner, so that they can be monitored and assessed, is a complex and challenging task [9] .

The integration of IA and automation system is seen in the larger scope of integration of intelligent devices that can interact and cooperate. Hence, from a technical point of view, when it comes to interaction some criteria from cooperative devices [10] could be adopted. These technical criteria include resource utilization, semantic description capabilities, inheritance/polymorphism, composition/orchestration, pluggability, service discovery, service direct/indirect device access, access to events, service life-cycle management, device management, security & privacy, and service monitoring.

ISO/IEC 25010 [6] , formerly ISO/IEC 9126, is a set of standards that proposes several quality characteristics to be taken into consideration for the evaluation of a software product. Although such quality models focus on software, they are well-suited for CPS. The ISO/IEC 25010 model can easily be adapted to the subject matter of this work, which investigates how to assess the integration of industrial (software) agents with low-level automation systems. While this constitutes an objective list of criteria that can be applied to assess a product, one has to also consider the actual usage of the product. To that end, ISO/IEC 25010 defines a *Quality in Use* model listing additional characteristics.

The Foundation for Intelligent Physical Agents (FIPA) defines specifications for the development of heterogeneous multi-agent systems. Among the several specifications, it addresses agent integration with software systems [11] , e.g., defining ontology and semantics, but lacks the establishment of specifications addressing the agent-device integration and corresponding validation and testing requirements.

There are various standards available for specifying, developing, and deploying industrial automation systems [12] , but when it comes to validation and testing of them there are only a few approaches and concepts are provided. In the domain of Programmable Logic Controllers (PLC) the widely used IEC 61131 describes, in part 2, the requirements and related verification tests addressing functional and electromagnetic compatibility issues [13] . Moreover, the user association PLCopen provides guidelines for the certification IEC 611313 environments; it focuses mainly on the compliant programming systems of PLCs where three different compliance levels are defined (i.e., Base Level - BL, Conformity Level - CL, and Reusability Level - RL) [14] . The interoperability approach IEC 61850 for Intelligent Electronic Devices (IED) used in power systems provides, in part 10, guidelines for conformance testing [15] . A comparable approach has been chosen by IEC 60870–5-6 where conformance testing guidelines for telecontrol equipment is being described [16] .

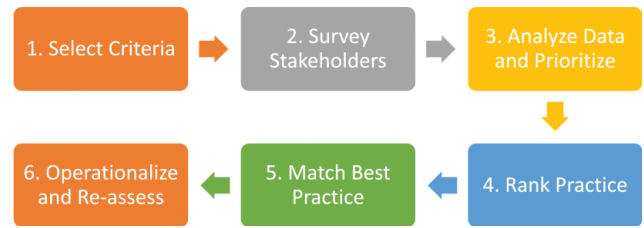

Figure 1. Methodology overview.

Summarizing, there are some approaches available from the industrial automation domain which look at validation and testing criteria but they have been mainly developed for conformity checks of a specific standard, e.g., for PLCs or for industrial communication as outlined above. Also, agent-based approaches like FIPA don't really provide concrete assessment criteria which are needed for evaluating the recommended practices for integrating IA and low-level automation.

## 3. PROPOSED METHODOLOGY

The main steps involved in the proposed methodology for assessing or choosing an appropriate practice integrating IA and low-level automation functions is shown in Figure 1 . The discrete steps of this methodology, ranging from understanding stakeholders' needs to realizing a system, can be used in varying sequences depending on the intended objective e.g.: (i) ranking the available practices or assessing the importance of characteristics for a specific use case can be realized via steps 1, 2, 3 and 4; (ii) selecting a practice from a list of previously ranked practices can be realized via steps 1, 2, 3, 5 and optionally 6.

*Step 1 – Select Criteria:* A key step of the process is to select the criteria upon which the evaluations can be done. Such criteria ought to consider the largest stakeholder group as well as the affected technical and business processes. Therefore, existing work carried out in research or standards organizations discussed previously in section 2 is of relevance. An example standard providing such list of criteria, and which is followed in this work, is ISO/IEC 25010. While the high-level criteria from such standards might be a good start, we might need to repeat the process with a more narrowed focus on technical aspects, e.g., if performance efficiency is a major issue, we have to see which of its sub-characteristics emerge in step 3.

*Step 2 – Survey Stakeholders:* Having the criteria defined and contextualized, the next step is to acquire feedback from the stakeholders. To do so, a survey can be conducted, where the stakeholders can provide their feedback in a quantitative form e.g., a Likert scale [17] . A quantitative assessment linked to the characteristics enables their statistical assessment.

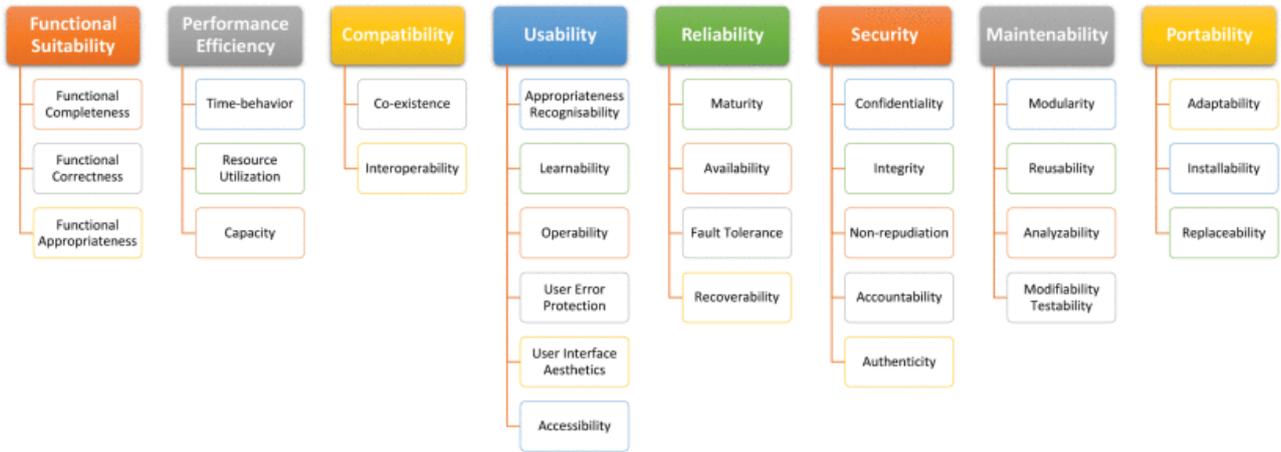

Figure 2. Overview of ISO/IEC 25010 product quality model characteristics.

*Step 3 – Analyze Data and Prioritize:* This step involves the analysis of quantified survey data, resulting in the attribution of values to the different (sub-)characteristics. The statistical analysis of the survey, reveals the views of the stakeholders, and provides insights on what is important and should be prioritized when selecting the approach to follow. For instance, from the quantitative values given to each characteristic, weights can be calculated that show the impact of that specific characteristic of the product quality model.

*Step 4 – Rank Practice:* At this step based on the empirical data collected by the survey and its analysis, each candidate practice can be assessed for suitability. This is typically done when one wants to rank a previously unranked practice or wants to have a view of what is important for a specific usecase, without the actual selection of a practice.

*Step 5 – Match Best Practice:* Assuming that there is a list of practices with quantitative scoring, one can do a similarity search and see which one fits best with the prioritized set of criteria from step 3. This step can help with the decisionmaking process of which practice to follow and implement.

*Step 6 – Operationalize and Re-assess:* Although most of the earlier steps help matching needs to a best practice prior to implementation, there is a need to assess how well an implemented practice performs in-use . Similar best practices may result in different user experience and satisfaction once operationalized, i.e., deployed and operated in the intentioned context. This step allows for an additional assessment considering the product use, which could provide additional operational insights for stakeholders. Assessment at this stage could help fine-tune the strategy, which may lead to enhancing the current practice or even replacing it with a better one. More quantitative data could enable a better understanding of practices and their suitability for a specific task.

## 4. ASSESSMENT CRITERIA

The first step according to the discussed methodology in section 3 is to select the criteria. The basis for this work is on the ISO/IEC 25010 product quality model. We have adapted the description of each of the characteristics in this model to make them more relevant to systems integrating IA and low-level automation and control. The close matching of the criteria and their interpretation to the domain is seen as important, as it will also help with the subsequent steps of stakeholder interaction, analysis, and prioritization.

### A. Assessment Scale

An important aspect of an assessment is to quantitatively assign a value to each criterion to highlight the relative importance of the characteristics. Such quantitative assessment may provide a rule of thumb, and enable statistical approaches to be used upon the criteria. For instance, the quantitative value given for each criterion could be in a 5-level Likert scale [17] , signifying how important the criterion is. As such, in a statement posed as *"Characteristic x is important for the integration of agents and low-level automation functions"* , the stakeholder would rate it as: (1) strongly disagree, (2) disagree, (3) neither agree or disagree (neutral), (4) agree, and (5) strongly agree. We have to explicitly point out that this is only just one of the potential approaches to quantify the impact of the respective characteristics. We have used this approach to perform a small survey of practitioners, which is further discussed in section 5.

### B. Quality Criteria

The characteristics as defined in ISO/IEC 25010 are being used in this work as the primary criteria for the assessment. ISO/IEC 25010 also lists sub-characteristics for each characteristic, which can have measurable quality attributes that can be quantifiably assessed over a scale, as discussed in subsection 4-A. We now put these characteristics and subcharacteristics of ISO/IEC 25010 [6] (an overview of which is shown in Figure 2 ), in the context of integrating industrial agents and low level automation and control:

*1) Functional suitability:* refers to the degree *"to which a product or system provides functions that meet stated and implied needs when used under specified conditions"* [6] .

The functionality of the interface integrating IA and low-level automation functions is a critical quality factor to ensure proper operation. Sub-characteristics include *completeness*, *correctness*, and *appropriateness*. The functional completeness of an interface is related to the level to which the set of functions provided by the interface covers all the specified tasks and user objectives. Functional correctness is the capability of the interface to produce correct results with respect to the specified behavior. At last, the functional appropriateness expresses the degree to which the functions facilitate the accomplishment of specified tasks and objectives.

*2) Performance efficiency:* refers to *"the performance relative to the amount of resources used under stated conditions"* [6]. Sub-characteristics include *time behavior*, *resource utilization*, and *capacity*. When discussing particularly the connectivity between agents and automation controllers like PLCs it becomes obvious that three aspects influence the overall system efficiency: the agent, the connection itself, and the automation device. This efficiency is then a function of the quality of the software but also of the supporting hardware platforms. Generally, the software implementation of the agent platform restricts their performance to soft realtime applications while the hard real-time layer is handled on the controller side. The quality of the integration is paramount here as it determines whether both sides will receive the right information at the right time within their independent control cycles. The stability of the connection (low jitter) needs to be balanced with the need for sheer throughput. Traditional measurements include assessing the distribution of the Round Trip Time (RTT) of the messages between agents and controllers and the estimation of the CPU load and memory footprint on both ends of the interaction. These are particularly difficult to estimate for large and complex agentbased system, for the general case, since the quality of the network infrastructure and the other support computational platforms greatly vary in capabilities and load.

*3) Compatibility:* refers to *"a product, system or component can exchange information with other products, systems or components, and/or perform its required functions, while sharing the same hardware or software environment"* [6]. ISO/IEC 25010 notes *co-existence* and *interoperability* as the two sub-characteristics of compatibility. For practices where IA are coupled with low-level automation functions, compatibility between components is applicable at two distinct levels. The first level is the boundary between agents and low-level control, where compatibility refers to agent's ability to seamlessly work with any variations of low-level automation functions, and vice versa. The second level concerns itself with the boundary between two or more coupled sub-systems, each containing agents and low-level control. In both cases, co-existence relates to the ability of components to run independently without affecting other components. Interoperability relates to ensuring homogeneity in data exchange format, protocols, and interfaces.

*4) Usability:* refers to the degree *"to which a product or system can be used by specified users to achieve specified goals with effectiveness, efficiency and satisfaction in a specified context of use"* [6]. Key sub-characteristics that provide insights to how well this is done are *appropriateness recognizability*, *learnability*, *operability*, *user error protection*, *user interface aesthetics*, and *accessibility*. In the specific context, usability is seen as partially relevant. For instance, its sub-criterion of interface aesthetics and accessibility, are not seen as critical, since usually there are no user interfaces that govern the way the integration of agents and low-level systems is done. However, other aspects, e.g., operability or learnability, might be more relevant if there is indirect user interaction. Overall though, since most agent related practices that integrate low-level automation functions operate on the background and usually as part of larger applications, but not directly with end-users, this criterion is seen as partly only relevant.

*5) Reliability:* refers to the degree *"to which a system, product or component performs specified functions under specified conditions for a specified period of time"* [6]. In industrial environments, interfaces between IA and low-level automation need to be reliable, with a level that increases with the criticality of the application. Sub-characteristics of reliability include *maturity*, *availability*, *fault tolerance*, and *recoverability*. Maturity is the capability of the interface to avoid failures, as a result of faults in the interface, and can be expressed using measures such as Mean Time To Failure (MTTF). Availability represents the fraction of time a system is operational, being desirable that high availability interfaces can be used. However, a high availability interface may fail, which may require the capacity of the system to continue operating properly despite the failure, which is expressed by the fault-tolerance sub-characteristic. In fact, an interface may fail due to several reasons, ranging from the incorrect design and implementation to the effects of the environment. Good fault-tolerant interface design requires the study of possible failures and the proper response to failures. Additionally, the recoverability sub-characteristic expresses the capability to mitigate the effects of an interface fault by recovering its performance and functionality as fast as possible.

*6) Security:* refers *"to which a product or system protects information and data so that persons or other products or systems have the degree of data access appropriate to their types and levels of authorization"* [6]. A typical system integrating IA with low-level control would be highly distributed and may have software interfaces to systems outside of the factory boundary. The three pillars of security are *confidentiality*, *integrity*, and *availability*. ISO/IEC 25010 captures the first two as security sub-characteristics while availability is captured as a sub-characteristic of reliability. In addition to these two, the standard also names non-repudiation, accountability, and authenticity as further security sub-characteristics. Confidentiality relates to the strength of encryption and access control across both distributed PLCs as well as the system and other external systems. Integrity relates to protecting exchanged data over these channels from being corrupted. Non-repudiation, accountability, and authenticity allow for monitoring intentional misuse by known or unknown users or systems though immutable system logs, digital signatures or authentication protocols. These aspects are becoming increasingly relevant as agents could be implemented by different stakeholders and aspects such as verification, confidentiality and availability are becoming increasingly important in industrial settings.

*7) Maintainability:* refers to the degree *"of effectiveness and efficiency with which a product or system can be modified to improve it, correct it or adapt it to changes in environment, and in requirements"* [6] . This aspect is very important to address also on IA and low-level automation side in industrial environments since products and therefore also the corresponding production systems and equipment might change over the total lifetime. Therefore, the automation systems (i.e., high-level IA and low-level automation) need to be designed in such a way that they can be adapted quickly to cope with new customer requirements and needs. According to ISO/IEC 25010 the most important sub-characteristics related to maintainability
are *modularity* , *reusability* , *analysability* , *modifiability* , and *testability* , all of which are important to software systems, including agent-based CPS that will need to satisfy them.

*8) Portability:* refers to the degree of *"effectiveness and efficiency with which a system, product or component can be transferred from one hardware, software or other operational or usage environment to another"* [6] . This, for instance, implies that the agent aspects should be portable and interact similarly if the hardware gets exchanged, or if part of the agent solution gets revised (e.g., a new agent in a multi-agent system), it should again execute similarly. Sub-characteristics for evaluation include *adaptability* , *installability* and *replaceability* .
Hence, any adjustment to software, e.g., agent or associated hardware part, should not affect the operational aspects, installations can be seamlessly done and components of the practice should be replaceable without affecting its operations.

**C. In-use Criteria**
While all of the criteria in subsection 4-B pose an objective list that can be applied to the product itself, one has to consider also the actual usage of the product, once it is operational. To that sense ISO/IEC 25010 also defines additional characteristics and which can provide insights once the implementation is finished and the practice is operationalized. Such insights can offer valuable feedback which can have two-fold use: (i) to re-assess the selected practice, and (ii) rank the practices by considering the operational aspects of it in selected deployment environments. The quality in-use characteristics which need to be considered are [6] :

*1) Effectiveness:* which captures the accuracy and completeness that the specific practice helps users achieve their goals. In industrial environments, effectiveness is of paramount importance.

*2) Efficiency:* which captures the utilized resources in relation to the accuracy and completeness that the goals are achieved. Efficiency is another highly relevant characteristic sought for operational industrial systems.

*3) Satisfaction:* which captures the user needs satisfaction. This is achieved via its sub-characteristics, i.e., *usefulness* , *trust* , *pleasure* , and *comfort* . In the agent integration with lowlevel automation functions, there has to be trust to the solutions itself and be useful, however, since this is a machine-based interaction other aspects like comfort may be less relevant.

*4) Freedom from risk:* which captures the mitigation of risks. Sub-characteristics include *economic* , *health & safety* , and *environmental risk* mitigation. Economic risks are to be considered, but the context of the agent-based solution, such as use in critical infrastructure, may elevate potential risks related to safety, or cascaded impacts of failure, to have equal or more importance.

*5) Context coverage:* captures the overall context aspects in which the product or system operates as well as beyond the initial explicitly identified context. Sub-criteria are *context completeness* and *flexibility* which capture again context-specific aspects, e.g., effectiveness, efficiency, freedom from risk, and satisfaction.

As it can be seen such characteristics are more qualitative, and highly depend on the operational environment and may be difficult to assess.

## 5. SURVEY AND DISCUSSION

As an example of utilizing the proposed methodology, a survey among experts in the domain of integrating industrial agents and low-level automation functions, was carried out. These experts are active in the *IEEE P2660.1* working group [4] . The aim of the survey is to get an initial indication into which of the overall characteristics and sub-characteristics of ISO/IEC 25010 are seen as important for this niche area. Hence, we followed steps of 1 to 4 of the methodology illustrated in Figure 1 , not to evaluate a specific practice, but to get an indication of what they consider as important.

The survey asked the experts to assess it in a 5-level Likert scale (as discussed in section 4 ), if they agree that the specific sub-characteristic is important. In total 17 industrial agent experts have filled in the survey. The results are shown in Table I . The right y axis in Table I shows the percentage of positive answers (that were graded agreement or strong agreement). As it can be seen the overwhelming majority of the items range from agreement to strong agreement. For instance 100% agree that *testability* is important, while for user interface aesthetics only 35% agree, with 41% disagreeing and 24% being neutral.

The skew towards agreement, and especially when investigating the positive agreement on the top-rated sub-characteristics, also reveals the most important aspects to pay attention to. These come as no surprise, considering the subject matter of this work which is industrial automation.

Table I. EXPERT SURVEY ON SUB-CHARACTERISTICS IMPORTANCE.

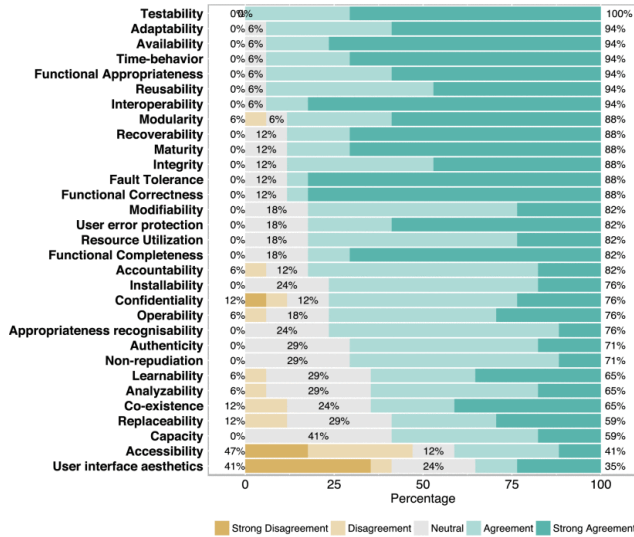

The high positive agreement to maturity (88%), availability (94%), fault tolerance (88%), and recoverability (88%), all of which are sub-characteristics to reliability, let us consider that this characteristic is highly important for industrial automation. Similarly, usability aspects, with some characteristics featuring high disagreement scoring, are not seen as much important since integration is done in the background and there is usually no significant interaction with end-users.

Some sub-characteristics of security are surprisingly low (in contrast to the rest in this survey), but this may be justified that up to now agent integration in automation functions was done to isolated environments (e.g., within the device or an attached host), where protection from malicious entities was not given much attention. However, as we move towards networked and collaborative infrastructures this is expected to change.

As demonstrated, the proposed approach can be applied and can lead to some insights. However, there are also several limitations when considering it. As it can be seen, the integration of industrial agents and low-level automation functions is approached from the product view, which captures several angles, that are attempted to be quantified. Such quantification, e.g., in Likert scale has its own limitations, e.g., the risk of leniency and severity errors [18] . In the example survey, shown in Table I , a view is expressed on the basis of the scores given by the respondents. These are IA integration experts, and although some bias may be present, the survey reflects the opinions of a specialized IA expert group. Hence, when most aspects are rated as important (as seen by the right skew of scores), these correspond to the common view among the experts. With larger and more diverse samples, a more diligent statistical analysis should be considered, e.g., Structural Equation Modeling (SEM).

The characteristics and sub-characteristics described in section 4 do not sufficiently capture in-depth the technical aspects of the approach, and which may pose the differentiating factor. The selection of a best practice for the integration of IA and low-level automation functions is very specialized, and requirements, as well as technologies, play a significant role when considering their impact on the acceptance of industrial agents as the statistical analysis of a recent survey suggests [5] . Similarly, operational aspects, i.e., the in-use criteria discussed in subsection 4-C , need to be more in-depth assessed as they also play a role in industrial agent acceptance, e.g., the cost is a significant factor for the decision makers [5] .

# 6. CONCLUSIONS

A methodology for assessing and selecting practices that integrate IA and low-level control in industrial automation systems is presented. The proposed methodology can be used in several ways - for ranking a set of available practices based on the quality requirements for a project, for identifying Powered by TCPDF (www.tcpdf.org) the best-fit practice for the project, and to assess and finetune an already operationalized practice. The methodology is supported by a robust set of quality criteria described in ISO/IEC 25010 software product quality model, which has been adapted for use in industrial automation systems in this paper. To exemplify the usage, a short survey of industrial automation experts involved in the IEEE P2660.1 working group was carried out, which provide some insights on what qualities are highly relevant. Future research directions to this work include expanding the criteria to measurable quality attributes, a wider empirical validation of the quality criteria, and studying the usage-specific strengths and benefits of the various practices for integrating and low-level control.